\documentclass[conference]{IEEEtran}
\usepackage[utf8]{inputenc}
\usepackage[T1]{fontenc}
\usepackage{graphicx}
\usepackage{amsmath}
\usepackage{amssymb}
\usepackage{booktabs}
\usepackage{array}
\usepackage{url}
\usepackage{hyperref}
\hypersetup{colorlinks=true,linkcolor=blue,citecolor=blue,urlcolor=blue}

\begin{document}

\title{Progressive Crystallization: Turning Agent Exploration into\\Deterministic, Lower-Cost Workflows in Production}

\author{\IEEEauthorblockN{Arun Malik}
\IEEEauthorblockA{Microsoft Azure Networking\\
Email: arunma@microsoft.com \quad ORCID: 0009-0005-6650-6711}}

\maketitle

\begin{abstract}
AI agents deployed for IT operations are typically permanent cost centers:
every execution re-invokes full LLM inference, even for problems the agent has
already solved many times. We present \emph{progressive crystallization}, a
lifecycle that treats agent exploration as a discovery mechanism rather than a
permanent execution model. We define an execution-type taxonomy spanning three
points, from fully agent-orchestrated (stochastic, expensive) through hybrid to
fully deterministic (zero-token, reproducible), and an evidence-based promotion
mechanism that converts an agent's repeatedly validated behavior into
progressively cheaper and more deterministic workflows, with automatic demotion
when a promoted workflow regresses. This makes an agentic platform cheaper,
faster, and safer over time without human rewriting of agent-discovered patterns.
In a production agentic system for cloud network operations handling tens of
thousands of incidents per month, the share of executions served by deterministic
workflows rose from zero to 45 percent over eight months, per-incident agent cost
fell by more than 70 percent while incident volume doubled, and safety properties
improved monotonically across promotions because determinism increases
reproducibility and auditability. We describe the taxonomy, the promotion and
demotion criteria, the trace-extraction method, an economic model, and the safety
argument, and we discuss limitations and threats to validity.
\end{abstract}


\section{Introduction}
Large language model (LLM) agents are increasingly deployed for IT operations
tasks such as incident triage, root-cause analysis, and automated
remediation~\cite{zhang2025aiops,malik2026air}. An agent observes system state
through tool calls, reasons about the problem, and executes actions in a loop.
This flexibility is exactly what lets an agent handle novel incidents, but it
also makes the agent a permanent cost center. Every execution re-invokes full
model inference, so the same recurring incident consumes thousands of tokens each
time it recurs, the investigation path is non-deterministic and hard to
reproduce, and total cost scales linearly with incident volume.

The core waste is that a successful agent investigation is discarded. When an
agent resolves an incident through a novel path, that knowledge is lost; the next
occurrence of the same failure re-discovers the solution from scratch, at the
same token cost and with a possibly different and inferior result. Existing
options do not close this gap. Traditional workflow engines cannot handle novel
scenarios and require an engineering sprint per new automation. Unconstrained
agents never get cheaper. Fine-tuning yields a smaller probabilistic model rather
than a deterministic workflow, and recorded-macro or robotic-process-automation
approaches capture surface actions, not the reasoning or the data flow, and are
brittle to environment change.

We argue that agent exploration should be a \emph{discovery mechanism}, not a
permanent execution model. Behavior that an agent discovers and validates through
repeated successful execution can be systematically converted into deterministic
workflows that require zero LLM tokens to run, while the agent layer remains
available for genuinely novel problems. We call this lifecycle \emph{progressive
crystallization}. This paper makes four contributions:
\begin{itemize}
  \item an \textbf{execution-type taxonomy} that places operational workflows on
  a spectrum from fully agent-orchestrated to fully deterministic
  (Section~\ref{sec:taxonomy});
  \item a \textbf{promotion and demotion lifecycle} that advances workflows down
  the spectrum based on accumulated evidence, and reverts them on regression
  (Section~\ref{sec:lifecycle});
  \item an \textbf{economic model} showing that platform inference cost decreases
  over time even as automation volume grows (Section~\ref{sec:economics}); and
  \item a \textbf{safety-monotonicity argument} that each promotion preserves or
  improves safety, supported by production evidence from a cloud-network agentic
  system (Sections~\ref{sec:safety}--\ref{sec:eval}).
\end{itemize}

\section{Related Work}
LLM agents that reason and act through tools were popularized by
ReAct~\cite{yao2023react} and Toolformer~\cite{schick2023toolformer}, and the
design space is surveyed in~\cite{xi2023agents,wooldridge2009multiagent}. Applying
these agents to operations is the focus of AIOps~\cite{notaro2021aiops} and its
LLM-era successors~\cite{zhang2025aiops,malik2026air}. Prior work largely treats
the agent as the permanent execution engine. Cost-reduction efforts such as
FrugalGPT~\cite{chen2023frugalgpt} lower per-call cost through model cascades and
routing, but the system remains probabilistic and never eliminates inference for
solved problems. Our contribution is orthogonal and complementary: rather than
making each inference cheaper, we remove inference entirely for work that has
been proven, by extracting deterministic workflows from execution traces. The
extraction step draws on process mining~\cite{vanderaalst2016procmining}, which
recovers process models from event logs; here the event logs are agent execution
traces and the recovered models are executable playbooks.

\section{Execution-Type Taxonomy}
\label{sec:taxonomy}
\begin{figure*}[t]
\centering
\includegraphics[width=0.92\textwidth]{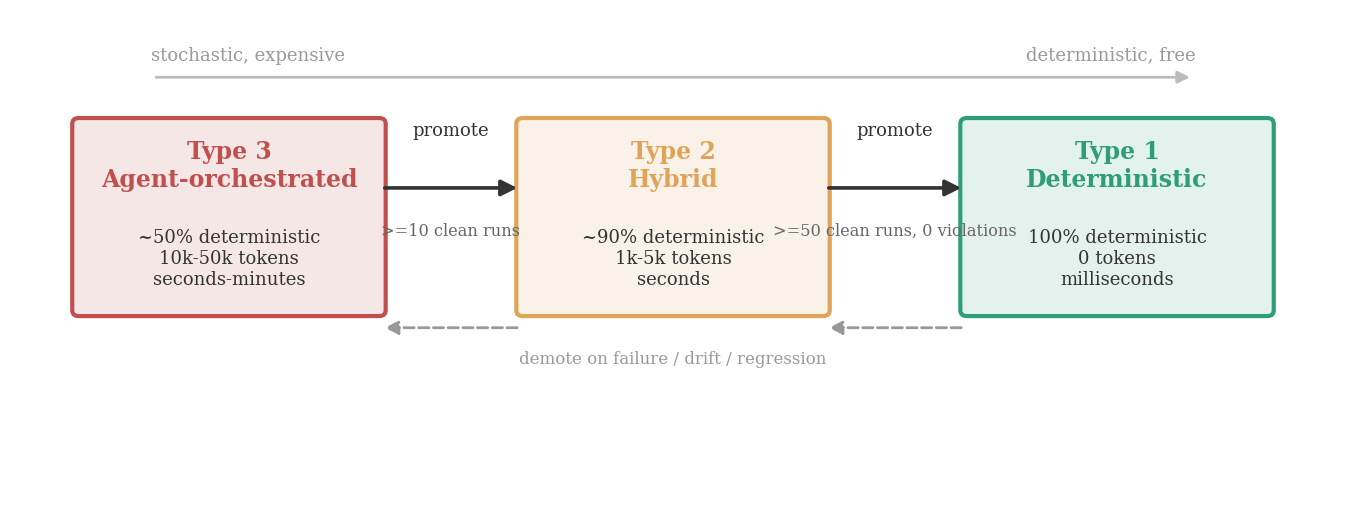}
\caption{The execution-type spectrum and crystallization lifecycle. Workflows are
promoted rightward, toward determinism and zero token cost, only on accumulated
evidence, and are demoted leftward automatically on failure, drift, or test
regression.}
\label{fig:spectrum}
\end{figure*}

We define three execution types for operational playbooks, forming a spectrum
from fully stochastic to fully deterministic (Fig.~\ref{fig:spectrum},
Table~\ref{tab:types}).

\begin{table}[t]
\caption{Execution-type taxonomy.}
\label{tab:types}
\centering
\small
\begin{tabular}{@{}p{1.4cm}p{1.95cm}p{1.3cm}p{1.5cm}@{}}
\toprule
\textbf{Type} & \textbf{Execution} & \textbf{Determ.} & \textbf{Token cost} \\
\midrule
Type 3: Agent-orchestrated & Sub-agents reason within bounded scope; deterministic checkpoints and HITL gates & $\sim$50\% & High (10k--50k) \\
Type 2: Hybrid & Structured steps; specific stages call the LLM for interpretation or classification only & $\sim$90\% & Low (1k--5k) \\
Type 1: Deterministic & Pre-coded logic, conditionals, and typed API calls; no LLM at runtime & 100\% & Zero \\
\bottomrule
\end{tabular}
\end{table}

A \textbf{Type~3} (agent-orchestrated) playbook lets the agent investigate freely
within a bounded scope, combining autonomous read operations with deterministic
checkpoints and human-in-the-loop (HITL) approval for write operations. It is the
entry point for any novel incident. A \textbf{Type~2} (hybrid) playbook has a
fixed step structure in which only specific stages invoke the LLM, and only for
interpretation, classification, or summarization; every action is typed and
schema-validated, so the LLM decides \emph{understanding}, not \emph{what to do}.
A \textbf{Type~1} (deterministic) playbook is pre-coded logic that runs with the
same guarantees as traditional workflow automation, at zero token cost and full
reproducibility. The three types are not separate systems; they are lifecycle
stages of the same discovered behavior.

\section{The Crystallization Lifecycle}
\label{sec:lifecycle}
Crystallization advances a playbook from a higher (more stochastic) execution
type to a lower one as evidence accumulates.

\textbf{Stage 1: discovery.} A novel incident with no matching playbook triggers a
Type~3 execution. The agent investigates using available tools; read operations
run autonomously and write operations pause for approval. The complete execution
trace is recorded.

\textbf{Stage 2: capture.} On verified successful resolution, the successful path
is extracted into a reusable Type~3 template. The extraction algorithm parses the
trace into an ordered list of tool calls, detects the branch conditions the agent
acted on, infers input and output schemas per step, builds a directed acyclic
graph of tool dependencies, parameterizes instance-specific values such as device
identifiers and timestamps, and marks human-approval points as explicit gates.

\textbf{Stage 3: promotion to hybrid.} After repeated successful runs of the
template, trace analysis identifies steps where the LLM consistently produces the
same classification (replaced by a deterministic rule) and steps where reasoning
varies but the outcome is stable (replaced by a scoped, single-purpose prompt).
Acceptance tests are generated automatically from the successful traces, and the
candidate Type~2 playbook must pass them.

\textbf{Stage 4: promotion to deterministic.} After further successful hybrid
runs without LLM disagreement, remaining LLM steps whose output is drawn from a
finite set, or whose decision boundary can be expressed as a rule, are replaced
with deterministic equivalents. The final Type~1 playbook needs no runtime tokens
and continues to be validated by the Stage~3 acceptance tests.

\begin{table}[t]
\caption{Evidence-based promotion criteria (defaults; configurable).}
\label{tab:promotion}
\centering
\small
\begin{tabular}{@{}p{1.6cm}p{6.0cm}@{}}
\toprule
\textbf{Transition} & \textbf{Requirements} \\
\midrule
Type 3 $\rightarrow$ 2 & $\geq$10 successful runs; zero safety violations; $\geq$90\% of runs produce the same action sequence; all auto-generated acceptance tests pass; no human override in the recent window \\
Type 2 $\rightarrow$ 1 & $\geq$50 successful hybrid runs; LLM classification consistency $\geq$99\%; the deterministic rule covers all observed input variation; full regression suite passes without the LLM; human review of the deterministic logic \\
\bottomrule
\end{tabular}
\end{table}

Promotion is gated by the criteria in Table~\ref{tab:promotion}. Crucially,
autonomy is attached to the specific playbook class and action type, based on
its evidence, rather than to the capability of the underlying model. A more
capable model does not automatically earn more autonomy; a track record does.

\section{Economic Model}
\label{sec:economics}
\begin{figure}[t]
\centering
\includegraphics[width=0.86\columnwidth]{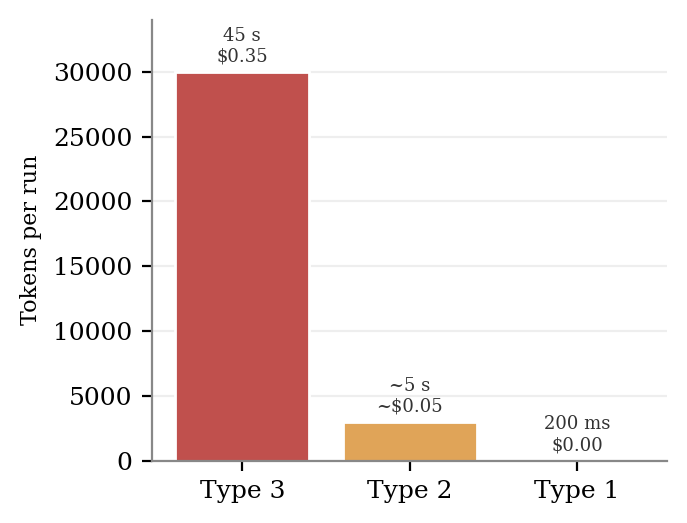}
\caption{Cost per run collapses across execution types. Latency and per-incident
dollar cost are annotated above each bar.}
\label{fig:cost}
\end{figure}

Each execution type has a characteristic cost (Fig.~\ref{fig:cost}). A Type~3 run
consumes roughly 10{,}000 to 50{,}000 tokens and completes in seconds to minutes;
a Type~2 run consumes on the order of 1{,}000 to 5{,}000 tokens; a Type~1 run
consumes none and completes in milliseconds. The key property is not the cost of
any single run but the shift in the \emph{mix} over time. As recurring patterns
crystallize, the platform routes each incoming request to the lowest-cost type
available for that pattern, so the fraction of Type~1 executions grows and total
inference cost falls even as automation volume rises. The agent layer becomes a
discovery mechanism whose cost is amortized across all future deterministic
executions of what it discovers, rather than a cost incurred on every occurrence.

\section{Safety Monotonicity}
\label{sec:safety}
A natural concern is whether making a workflow cheaper makes it less safe. We
argue the opposite: crystallization preserves or improves every safety property
we track (Table~\ref{tab:safety}). Auditability is unchanged, since all three
types record full traces. Reproducibility rises monotonically, from roughly 50
percent for Type~3 to 100 percent for Type~1. Blast-radius control moves from
runtime HITL gates (Type~3) to schema validation (Type~2) to statically
verifiable deterministic logic (Type~1), which is stronger because it can be
checked before execution rather than caught during it. Compliance shifts from
conditional (depends on a human being present) to built-in. In short, a
deterministic playbook is easier to verify, reproduce, and audit than the agent
run it was distilled from, so safety and cost improve together rather than
trading off.

\begin{table}[t]
\caption{Safety properties across execution types.}
\label{tab:safety}
\centering
\small
\begin{tabular}{@{}p{2.3cm}p{1.5cm}p{1.4cm}p{1.4cm}@{}}
\toprule
\textbf{Property} & \textbf{Type 3} & \textbf{Type 2} & \textbf{Type 1} \\
\midrule
Auditability & Full trace & Full trace & Full trace \\
Reproducibility & $\sim$50\% & $\sim$90\% & 100\% \\
Blast-radius control & HITL gates & Schema validation & Deterministic logic \\
Compliance & Conditional & Built-in & Built-in \\
\bottomrule
\end{tabular}
\end{table}

\section{Demotion and Continuous Discovery}
\label{sec:demotion}
Crystallization is not one-way. Each promoted playbook is monitored, and a
circuit breaker demotes it to a higher execution type on execution failure,
safety violation, or acceptance-test regression. In production, a deterministic
playbook once broke after a firmware update changed a command's output format;
the deterministic parser failed, the system demoted the playbook to hybrid so the
LLM could handle the new format, and after a run of clean executions it was
re-promoted. This gives the platform the reliability of deterministic automation
for stable patterns and the adaptability of agents for change, without a human
deciding when to switch. Genuinely novel incidents always enter as Type~3, so the
discovery pipeline never closes.

\section{Production Evaluation}
\label{sec:eval}
We deployed progressive crystallization in a production agentic platform for
cloud network operations that resolves incidents across a large managed network
and handles tens of thousands of incidents per month~\cite{malik2026air}. We
report three observations.

\begin{figure}[t]
\centering
\includegraphics[width=0.9\columnwidth]{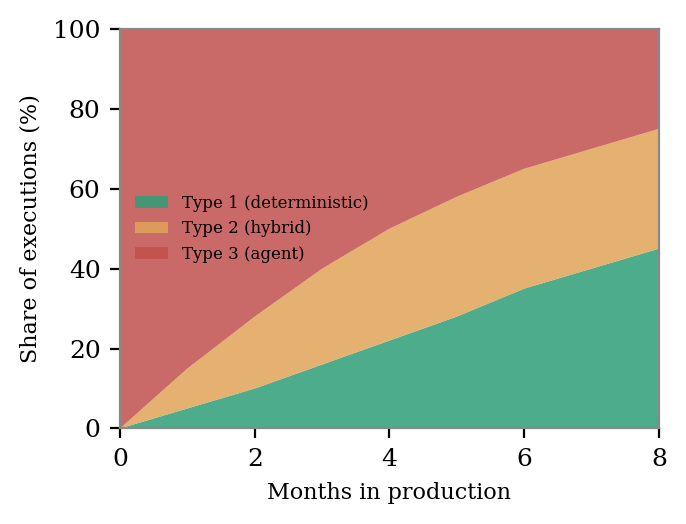}
\caption{Share of executions by type over eight months in production. Deterministic
workflows grow from zero to about 45 percent as patterns crystallize, while
agent-orchestrated executions fall to about a quarter.}
\label{fig:ratio}
\end{figure}

\textbf{The mix shifts toward determinism.} At launch, essentially all executions
were Type~3. Over eight months the share of deterministic (Type~1) executions rose
from zero to about 45 percent, with roughly 30 percent hybrid and 25 percent
agent-orchestrated (Fig.~\ref{fig:ratio}). The ratio of Type~1:2:3 executions is a
useful platform-maturity metric.

\textbf{Cost falls as volume rises.} Over the same period, per-incident agent cost
fell by more than 70 percent while incident volume roughly doubled. This is the
central economic claim of the paper realized in production: the platform got
cheaper as it did more, because it stopped paying for inference on work it had
already learned.

\textbf{Autonomy and quality held.} The platform resolves over 90 percent of
common incident categories autonomously, mean time to resolution fell from hours
to minutes, and the false-positive remediation rate stayed under 5 percent with
no customer-visible impact. Crystallization did not trade quality for cost; the
deterministic paths are the ones that had already proven reliable as agent runs.

\section{Discussion, Limitations, and Threats to Validity}
Results come from a single organization and operational domain, so specific
thresholds and ratios should be re-derived elsewhere; the lifecycle itself is
domain-agnostic. Crystallization assumes recurring patterns, so its benefit is
smaller in environments dominated by genuinely novel, one-off incidents, where
most executions will remain Type~3. The economic figures are platform-level
observations rather than a controlled comparison, and the eight-month window
reflects one maturity trajectory. Automatic promotion depends on the quality of
the acceptance tests generated from traces; a pattern that is under-observed can
be promoted prematurely, which is why demotion and human review of the final
deterministic logic are part of the design. Finally, extraction quality depends
on the richness of execution traces; sparse or poorly structured logging limits
what can be crystallized, echoing a known dependency in process mining.

\section{Conclusion}
Agent platforms do not have to be permanent cost centers. By treating agent
exploration as discovery and progressively crystallizing proven behavior into
deterministic, zero-token workflows, an agentic system can become cheaper,
faster, and safer over time without human rewriting. In production, the mix of
executions shifted from fully agent-orchestrated toward deterministic, cost fell
by more than 70 percent as volume doubled, and safety improved monotonically
because determinism buys reproducibility and verifiability. We believe the
Type~1:2:3 ratio is a practical maturity metric for any team operating agents at
scale, and that the discipline of promoting and demoting workflows on evidence is
a general pattern for making autonomy both affordable and trustworthy.

\bibliographystyle{IEEEtran}
\bibliography{references}

\end{document}